\documentclass{aastex}
\usepackage{spr-astr-addons}
\usepackage{url}

\RequirePackage{color}

\begin{document}

\title{Anisotropic dark energy stars}
\slugcomment{to appear in Astrophysics and Space Science}
\shorttitle{Anisotropic dark energy stars}
\shortauthors{Cristian R. Ghezzi}

\author{Cristian R. Ghezzi} 
\affil{Departamento de F\'{\i}sica,
Universidad Nacional del Sur, \\
Av. Alem 1253, 8000 Bah\'{i}a Blanca, Buenos Aires, Argentina. email: ghezzi@uns.edu.ar}

\begin{abstract}
A model of compact object coupled to  inhomogeneous anisotropic  dark energy is studied.  It is assumed a variable dark energy that suffers a phase transition at a critical density. 
 The anisotropic $\Lambda-$Tolman-Oppenheimer-Volkoff equations are integrated to know the structure of these objects. The anisotropy is concentrated on a thin shell where the phase transition takes place, while the rest of the star remains isotropic. The family of solutions obtained depends on the coupling parameter between the dark energy and the fermionic matter. 
The solutions share several features in common with the gravastar model.
There is a critical coupling parameter that gives non-singular black hole solutions. The mass-radius relations are studied as well as the internal structure of the compact objects. The hydrodynamic stability of the models is analyzed using a standard test from the mass-radius relation.
For each permissible value of the coupling parameter there is a maximum mass, so
the existence of black holes is unavoidable within this model.

\end{abstract}



\section{Introduction}
A star that consumed its nuclear fuel, if massive enough, will end its life as a compact object or a black hole.
The black holes described by the Einstein theory of gravity contain singularities.  However quantum effects must be taken into account at high curvature values, or short distances compared with the Planck length scale \citep{frolov}. The quantum effects can render an inner portion of the black hole as a de Sitter spacetime. This was considered by several researchers, see for example: \cite{gliner}; \cite{gliner2}; \cite{israel3}, and references therein. Recently, \cite{Nicolini08}, and \cite{Nicolini06}, obtained regular black hole solutions with an inner de Sitter portion through a non-commutative geometric approach to quantum gravity.  

Analogous concepts were extended to a set of models which replace the whole black hole region with a de Sitter spacetime \citep[see][]{chapline01,chapline03,mazur01,mazur04}). 
In fact, \cite{chapline01}, built a model of dark energy stars based on the analogy between a  superfluid condensate near its critical point with the neighborhood of an event horizon. The dark energy stars have a de Sitter spacetime matched with a Schwarzschild exterior spacetime. The surface of phase transition is closely located above the Schwarzschild radius.

The model of \cite{mazur01} (MM) is conceptually different from the dark energy stars, but possess several common features inspired in it.  Their model is called ``gravitational vacuum  stars'' ({\it gravastars}). They considered that independently of the matter that composes the gravitating object, as some critical limit is reached, the interior spacetime suffers a gravitational Bose-Einstein condensation \cite{mazur01,mazur04}. 
The general relativity is not valid in the zone of coexisting phases, but it is valid macroscopically out of the transition zone (\citeauthor{chapline01}; and \citeauthor{mazur01}). The net effect of the condensation is that the mean value of the vacuum stress-energy tensor changes from a nearly zero value to a non-zero value. The mean value of the vacuum stress-energy tensor has the form: $<T_{\mu \nu }>=<\rho _{vac}>g_{\mu \nu }$ \citep[see][]{weinberg2, dymnikova00}, which behaves like a cosmological constant term with $\Lambda =8\pi G c^{-2}<\rho _{vac}>$. This is equivalent to  dark energy of the cosmological constant type, and behaves like a fluid with an equation of state: $\hat{P}=-\rho\,c^2 $. The MM model is a static, spherical symmetric, five layer solution of the Einstein equations.  It was built cutting and pasting three different exact solutions to the Einstein field equations. It has an interior de Sitter spacetime, while the exterior is Schwarzschild.  The intermediate zone is a thick shell of matter that satisfies all the energy conditions. The three solutions are matched by two thin anisotropic layers with distributions of surface tension and  surface energy density, that violate the strong energy conditions.   The solution does not possess an event horizon, but has a compactness that is close to one (the black hole compactness).  

Several researchers have analyzed the gravastar solutions using semi-analytic or analytic methods. For example, 
\cite{faber} proposed a thick shell anisotropic gravastar model which has the advantage of smoothing out the large pressure jump present in the original MM model. \cite{dymnikova00} found new analytic solutions of the gravastar type, G-lumps, and $\Lambda $-black holes. Related new solutions of nonsingular black holes were also found by \cite{mbonye}. Their solutions  have an interior de Sitter  geometry containing matter. \cite{chirenti} found thick and thin shell gravastar solutions and studied their stability. \cite{chan} studied anisotropic phantom energy stars; while \cite{lobo} studied gravastar solutions with a quintessence-like equation of state. \cite{Nicolini06} obtained ``mini-gravastar"  and regular black hole solutions.

Due to the large compactness of the gravastars it could be difficult to distinguish a gravastar from a black hole. Some arguments against the existence of gravastars, that could be verifiable through observations, were given by \cite{narayan}. On the other hand, positive detection features must include high energy particles emitted from the surface of the dark energy stars, as pointed out by \cite{barbieri}.  The  model of dark energy stars, or gravastars, is an interesting model which could alleviate the black holes singularities and it is worth to explore its characteristics.

A fluid of the cosmological constant type has necessarily a constant density in the absence of matter or other fields \citep{dymnikova00}. The reason is that for a source of the type $T_{\mu \nu}=(8 \pi G)^{-1}\Lambda g_{\mu \nu}$, the conservation equation ${T^{\mu \nu}}_{;\nu}=0$, implies a constant $\Lambda$. Thus, in order to form a condensate the dark energy must be coupled to the matter.
On one hand, this can be achieved by a direct proportionality between the dark energy and matter.
  Another possibility is to consider a pure inhomogeneous anisotropic vacuum like term  \citep[see][]{mazur01,dymnikova00}. In this work these two possibilities are combined. Thus, a set of compact objects made of fermionic matter coupled to inhomogeneous, anisotropic, dark energy is studied in this paper. 

 The obtained solutions form a one-parameter family, where the parameter is the proportionality constant between the dark energy and matter.  The solutions converge on a model of the gravastar type as the parameter approaches a finite value. In addition, a new method is reported  for the numeric integration of the anisotropic Tolman-Oppenheimer-Volkoff equations with cosmological constant ($\Lambda$-TOV)  over the whole star. 

In the gravastar and dark energy star models the shell is located above the event horizon. The exact position of the shell can not be given as a boundary condition of the $\Lambda$-TOV equations (without relaxing some other boundary condition), because the coordinates of the Schwarzschild radius are determined by the full solution. In this work, the position of the thin shell is obtained self-consistently as a result of the numeric integration. The mass-radius relation and structure of the stable models are estimated and compared with the Tolman-Oppenheimer-Volkoff neutron stars.

The paper is organized as follows: in section \ref{sec2} the Einstein equations and the notation  are introduced. The $\Lambda$-TOV anisotropic equations are derived. In section \ref{EOS} the equation of state for the matter and dark energy are given. In this section, a surface tension at the shell is obtained as function of the anisotropy. The junction and boundary conditions for the $\Lambda$-TOV equations are discussed. In section \ref{numeric}, the numerical algorithm is explained. In section \ref{disc}, the results are discussed. The paper ends in  section \ref{remarks} with some final remarks.

\vspace{0.3cm}

\section{\label{sec2}Einstein equations}

The Einstein equations are:
\begin{equation}
R^{\mu \nu }-\frac{1}{2}g^{\mu \nu }R=\frac{8\pi G}{c^{4}}T^{\mu \nu }\,,
\end{equation}
where $R$ denotes the scalar curvature,  $R^{\mu \nu }$ is the Ricci tensor, and $T^{\mu \nu }$ is the energy momentum tensor.

Assuming spherical symmetry the line element in standard coordinates \citep{weinberg} is: 
\begin{eqnarray}
ds^{2}=-e^{\nu(r)}dt^{2}+e^{\lambda(r)}dr^{2}   \label{comline} 
+r^{2}(d\theta +\mathrm{sin}^{2}\theta d\phi ^{2})\,.
\end{eqnarray}
 The energy-momentum tensor is composed of matter with mass-energy density $\delta $ and  pressure $P$, plus  dark energy with density $\rho _{de}$, radial pressure $P_{de(r)}$, and tangential pressure $P_{de(t)}$: 
\begin{eqnarray}
&&T_{0}^{0}=\delta c^{2}+ \rho _{de}\,c^{2} \\
&&T_{1}^{1} = -\left(P+P_{de(r)}\right) \\
&&T_{2}^{2} = -\left(P+P_{de(t)}\right) = T_{3}^{3} \\
&&T_{1}^{0}=T_{0}^{1}=0\,
\end{eqnarray}

In terms of the (variable) cosmological constant the dark energy density is: $\rho _{de}=\Lambda \,c^{2}/8\pi G$. The energy density is written as the rest mass density plus the internal energy density of the gas: $\delta=\rho\,(1+\epsilon/c^2)$ (see \cite{shapiro}).

The dark energy radial pressure is proportional to the dark energy density: 
\begin{equation}
P_{de(r)}=-\rho _{de}\,\,c^{2}\,.
\end{equation}

The variable dark energy density is assumed to be proportional to the mass
density \mbox{$\rho _{de}=\alpha\,\rho $}, where $\alpha$ is a non-negative constant (see section \ref{EOS}). To sum up, the subindex $``de"$ (dark energy) will not be used from now on: $P_{de(r)}=P_{r}$ and $P_{de(t)}=P_{t}$, except for the dark energy density $\rho_{de}$.

The components of the Einstein equations are \citep[see][]{weinberg,ghezzi}: 
\begin{eqnarray}
&&T_{0}^{0}\,\,\mbox{:}\,\,8\pi (\delta +\rho _{de})/c^2=  \label{T00} \\ \nonumber
&&\hspace{0.8cm}e^{-\lambda }\left( \frac{\lambda ^{\prime }}{r}-\frac{1}{%
r^{2}}\right) +\frac{1}{r^{2}}
\end{eqnarray}

\begin{eqnarray}
&&T_{1}^{1}\,\,\mbox{:}\,\,8\pi (P+P_{r})/c^4=  \label{T11} \\   \nonumber
&&\hspace{0.8cm}e^{-\lambda }\left( \frac{\nu ^{\prime }}{r}+\frac{1}{r^{2}}%
\right) -\frac{1}{r^{2}}   
\end{eqnarray}

\begin{eqnarray}
&&T_{2}^{2}=T_{3}^{3}\,\,\mbox{:}\,\,8\pi (P+P_{t})/c^4=  \label{T22} \\   \nonumber
&&\hspace{0.8cm}e^{-\lambda }\left( \frac{\nu^{\prime \prime}}{2}-\frac{1}{4}\lambda^{\prime} \nu^{\prime}
+\frac{1}{4}(\nu^{\prime})^2+(\nu^{\prime}-\lambda^{\prime})/2r\right)
\end{eqnarray}

An integral of equation (\ref{T00}) is :

\begin{equation}
e^{-\lambda }=1-\frac{2mG}{rc^{2}}\,,  \label{grr}
\end{equation}

The mass-energy up to the radius $r'$ is: 
\begin{equation}
m(r')=4\pi \int_{0}^{r'}(\delta +\rho _{de})\,r^{2}dr,
\label{totalmass}
\end{equation}
 In the special case of a constant dark energy density: $m=m^{\prime}+\frac{1%
}{6}(\frac{\Lambda c^{2}}{G})r^{3}\,$, where $m^{\prime }$ is the integral (\ref{totalmass}) performed over the  mass-energy density of the matter alone.

Subtracting Eq. (\ref{T11}) from Eq. (\ref{T00}) it is 
\begin{equation}
\frac{e^{-\lambda }}{r}\frac{d(\lambda +\nu )}{dr}=-\frac{8\pi}{c^2} \left( \delta+P/c^2\right) \,.  \label{metrica0b}
\end{equation}
At the surface of the star the right side of the equation is zero, so $\lambda +\nu $
is independent of $r$. To get an asymptotic flat solution it should be 
$\lambda, \nu \rightarrow 0$ (so $\lambda +\nu \rightarrow 0$) as $r\rightarrow \infty $, thus: 
\begin{equation}
\lambda =-\nu \,\,\,\,\,\,\,\,\mathrm{for}\,\,\,\,r \ge r_{s},
\label{coefmetrica1}
\end{equation}
implies $g_{tt}=g_{rr}^{-1}$ for $r \ge r_s$, where $r_s$ is the surface radius.
It can be seen from equation (\ref{grr}) that $m(r)\rightarrow 0$, as $r \rightarrow 0$, in order to get a regular metric at the center. So, an integration constant in equation (\ref{totalmass}) was set to zero.
The equation (\ref{T11}) can be cast as:
\begin{equation}
\frac{1}{2}\nu^{'}=\frac{4 \pi (P+P_r) r^3/c^2+m G/c^4}{r(r-2mG/c^2)}\,.
\label{nuprime}
\end{equation}
\vspace{0.2cm}
\subsection{\label{hydro}Hydrostatic equilibrium structure equations}

The hydrostatic equilibrium equation can be obtained from the Einstein equations: 
\[
\hat{P}_{,r}+\frac{a_{,r}}{a}(\delta c^{2}+P)+\frac{2}{r}(P_{r}-P_{t})=0\,, 
\]
where the notation $a=e^{\nu },$ is used, and $\hat{P}=P+P_{r}$ is the total radial pressure. Rearranging the terms the anisotropic
$\Lambda$-Tolman-Oppenheimer-Volkov ($\Lambda$-TOV) equation is obtained: 
\begin{eqnarray}
&&\frac{d\hat{P}}{dr}=-(\delta c^{2}+P)\frac{\bigl(\frac{mG}{c^{2}}+\frac{4\pi G}{%
c^{4}}\hat{P} r^{3}\bigr)}{r\bigl(r-\frac{2mG}{c^{2}}\bigr)}  \nonumber \\  \label{TOV1}
&&\,\;\;\;\;\;\;+2\frac{\Delta P}{r},
\end{eqnarray}
where 
$$\Delta P=P_{t}-P_{r},$$
is the anisotropic term. This equation and its solutions were first studied by \cite{bowers}. Only for
completeness, it is possible to rewrite the equation above for matter and uncoupled dark
energy:
\begin{eqnarray*}
&&\frac{d\hat{P}}{dr}=2\frac{\Delta P}{r}+ \\
&&-(\delta c^{2}+P)\frac{\bigl(\frac{m^{\prime}G}{c^{2}}+\frac{4\pi G}{c^{4}}Pr^{3}-%
\frac{8\pi G}{3c^{2}}\rho _{de}r^{3}\bigr)}{r\bigl(r-\frac{2m^{\prime}G}{c^{2}}-\frac{%
8\pi G}{3c^{2}}\rho _{de}r^{3}\bigr)}\,,  \label{TOV2}
\end{eqnarray*}
or as function of the cosmological constant  $\Lambda =8\pi G\,\rho _{de}/c^{2}\,$:
\begin{eqnarray*}
&&\frac{d\hat{P}}{dr}=2\frac{\Delta P}{r}+ \\
&&-(\delta c^{2}+P)\frac{\bigl(\frac{m^{\prime}G}{c^{2}}+\frac{4\pi G}{c^{4}}Pr^{3}-%
\frac{1}{3}\Lambda r^{3}\bigr)}{r\bigl(r-\frac{2m^{\prime}G}{c^{2}}-\frac{1}{3}%
\Lambda r^{3}\bigr)}\,.
\end{eqnarray*}
This equation reduces to the well known Tolman-Oppenheimer-Volkov (TOV) 
equations when $\Delta P=0,$ and $\Lambda =0$ \cite[see][]{oppi}. In this paper, equation (\ref{TOV1}) is  numerically integrated. 

\section{\label{EOS}Equation of State}
The equations obtained above must be supplemented with equations of state for the matter and for the dark energy.

\subsection{Equation of state for the matter\label{seceos}}

 A gas of neutrons at zero temperature has been considered here. 
This has led to a direct comparison of the results
with the \cite{oppi} neutron star model. The energy per unit mass of the matter is given by  \citep{shapiro}: 
\begin{eqnarray}
\delta \,c^{2}=\frac{m_{n}^{4}c^{5}}{\hbar ^{3}}\frac{1}{8\pi ^{2}}\biggl[ %
x\,\sqrt{1+x^{2}}\,\bigl(1+2x^{2}\bigr)-     \\ \nonumber
\mathrm{ln}\bigl(x+\sqrt{1+x^{2}}\bigr) \biggr]\,,
\label{EOSe2}
\end{eqnarray}
 This gives the total energy of the gas, including the rest mass energy density $\rho$,  measured in [$\mathrm{erg\,cm}^{-3}$].
The Fermi $x$ parameter is: $x=p/m_{n}c$, where $p=(\frac{3h^{3}}{8\pi 
}n)^{-1/3}$ is the momentum of the particles, and $n$ is the number density of neutrons. It can be written: 
\begin{equation}
x=\biggl( \frac{\rho }{\rho _{0}}\biggr)^{1/3}\,,
\end{equation}
with: 
\begin{equation}
\rho _{0}=\frac{m_{n}^{4}c^{3}}{3\pi ^{2}\hbar ^{3}}=6.106\times
10^{15}\,\,\,\,[\mathrm{g\,cm}^{-3}]\,.
\end{equation}

The pressure is given by  \citep{shapiro}: 
\begin{eqnarray}
\label{eosn}
P=\frac{m_n^4 c^5}{\hbar^3} \frac{1}{8 \pi^2} \biggr[x\, \sqrt{1+x^2} \,%
\bigl(2 x^2/3-1\bigr) + \\
\mathrm{ln} \bigl(x+\sqrt{1+x^2}\bigr) \biggr]\,,
\end{eqnarray}
measured in [$\mathrm{dyn\, cm}^{-2}$]. 
In this work the matter is assumed isotropic.

\subsection{Equation of State for the dark energy}
 The dark energy is assumed to be proportional to the rest mass if its density is above a certain critical value $\rho_{c}$, i.e.: 
\begin{eqnarray}
\rho_{de}&=&\alpha\,\rho\,\,\,\,  \rm{for} \,\,\rho \geq \rho_{c}\\ \nonumber
& &0 \,\,\,\,\,\,\,\,\, \rm{for} \,\,\,\,\,\, \rho < \rho_{c} , 
\end{eqnarray}
where $\alpha $ is a non-negative proportionality constant (here called ``coupling parameter").  

 In this case the radial dark pressure is: 
\begin{eqnarray}
P_{r}&=&-\rho _{de}\,c^{2} \,\,\,\, \rm{for} \,\,\rho \geq \rho_{c}\\ \nonumber
& &0 \,\,\,\,\,\,\,\,\,\,\,\,\,\,\,\,\,\,\,\,\, \rm{for} \,\, \rho < \rho_{c} ,
\label{eos1}
\end{eqnarray}

Observe that the equation of state of the matter plus the dark energy  is analogous to the  MIT bag model for hadrons \citep[see][pags. 323-324]{glendenning}. The dark energy equation of state given above, corresponds to the bag constant contribution of the MIT bag model. The bag constant is taken as density dependent in this work. The MIT bag model is used in astrophysical models of hybrid stars with a deconfined phase of quarks at the core of the star. For isospin symmetric nuclear matter the quark-hadron phase transition must occur at a critical mass density of about two to five times the saturation nuclear density, although its certain value is unknown. But recently was considered that for isospin asymmetric nuclear matter and temperatures of several MeV the onset of the transition to quark matter could happen already around saturation nuclear density, or even smaller densities, depending on the bag constant. This do not contradict accelerator physics, mainly because dynamical timescales in collisions are very short ($10^{-23}$ s), compared to weak time processes ($10^{-6}-10^{-8}$ s), and much shorter than dynamical timescales in explosive events like supernova explosions, which are of the order of ms. So strangeness can be produced and maintained in weak equilibrium in  astrophysical enviroments  \citep{sagert}. This means that it is possible to consider a sub-saturation critical density for the deconfinement transition in a compact object.

The model studied here can be thought as an extrapolation of an hybrid star model, where the bag constant takes arbitrarily large values. In the spirit of the original gravastar model it is assumed that the vacuum
phase transition is induced (or enhanced) by a gravitational field condensation effect. As will be discussed below, this lead to solutions that interpolate between normal neutron stars and gravastars. 

According to this considerations, it is set the critical density: $\rho_{c}=2\times 10^{14}\,$g cm$^{-3}$.  
The dependence of the solutions on $\rho_c$ will not be discussed here, for simplicity, because the solutions are qualitatively similar but with lower maximum mass (if $\rho_c$ is larger).

The transition in the vacuum energy density is assumed as enhanced by the gravitational Bose-Einstein condensation \citep{chapline03}. 
As the negative pressure of the dark energy helps to maintain a larger mass compact object, the dark energy density will be assumed without upper bound in order to get the maximum masses allowed by the present model.
 
The tangential dark energy pressure and the radial dark energy pressure differ only at the shell of phase transition: $P_{r} \ne P_{t}$, at $\rho = \rho_{c}$. Thus, the dark energy is in general anisotropic at the shell of phase transition.
 The tangential pressure $P_{t}$ is  determined by the field equations. 

\begin{figure}[h]
\begin{center} 
\includegraphics[width=8cm]{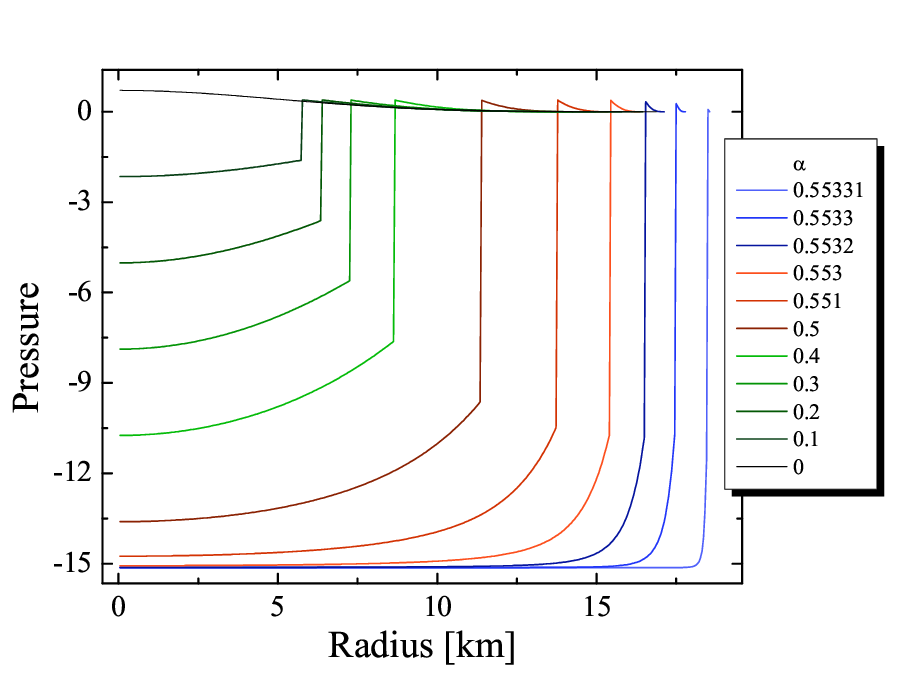} 
\end{center}
\caption{Pressure profile for each star. The pressure tends to a constant negative value inside the core as the critical parameter ($\alpha_c =0.5533$) is approached. The pressure is measured in  [${\rm g\,cm}^{-3}$] and normalized by the factor $10^{13}\,c^2$.}
\label{stab2}
\end{figure}

\subsubsection{\label{anisot}Anisotropy and surface tension}
There is a surface tension at the interface of phase transition. This is expected, in general, at the interface between two substances.  In the absence of matter, the equations for pure anisotropic, inhomogeneous, dark energy \citep{dymnikova00} are
obtained from equations (\ref{TOV1}) and (\ref{eos1}): 
\begin{eqnarray}
&&P_{r}=-\rho _{de}c^{2}\,,  \nonumber \\
&&P_{t}=P_{r}+\frac{r}{2}\frac{dP_{r}}{dr}\,.  \label{DETOV}
\end{eqnarray}
 A surface tension can be defined as:
\begin{equation} 
\sigma =(\Delta P)\,\delta r,  \label{sigma}
\end{equation}
measured in dyn/cm. Here $\delta r$ is the width of the shell, which is expected to be of Planckian thickness $l_{P} \sim 10^{-33}$ cm. With these definitions, equation (\ref{DETOV}) can be written as:
\begin{equation}
\left[ P_{r}\right] =2\sigma /r\,, \label{sigma2}
\end{equation}
where $\left[ P_{r}\right] =P_{r}^{+}-P_{r}^{-}$, is the jump in the dark radial
pressure,  $P_{r}^{\pm }$ is the radial pressure evaluated above/below the
shell, respectively, and $2\sigma /r\,$ is evaluated at the shell's radius $r.$ 
 Equation (\ref{sigma2}) is analogous to a Young-Laplace equation for the spherical interface between two substances \citep{landau}. The anisotropy (surface tension) compensates the large dark energy pressure gradient. 
From Eq. (\ref{sigma2}) above, it follows that a positive surface tension implies a larger radial dark pressure from above the thin shell \footnote{In fluid dynamics \citep{landau} the Eq. (\ref{sigma2}) is commonly written $-\left[ P_{r}\right] =2\sigma' /r\,$, with $\sigma'=-\sigma$. Thus, what is called a positive surface tension $\sigma$ here is a negative surface tension in the textbooks, and vice versa.}. There is a net radial pressure compressing the shell. In this case, the tangential dark pressure is larger than the radial one on the shell (see Eq. \ref{sigma}). On the contrary, a negative surface tension, gives a net radial dark pressure directed outwardly, and a tangential dark pressure lower than the radial one. The model analyzed in this paper has positive surface tension.

The surface tension  is not directly calculated (and it is not necessary) in the algorithm, but it could be found with an indirect calculation. It can be observed that it remains finite. This is inferred  {\it a posteriori} from the numeric results,  using  the Israel-Lanczos \citep{israel2} thin shell junction conditions. As observed by  \cite{mazur01}, the coefficient $g_{tt}$ of the metric is continuous, but its first derivative and the metric coefficient $g_{rr}$ are in general discontinuous, due to the discontinuity in the equation of state. The outwardly directed normal vector to the interface is ${\bf n}=(\sqrt{g_{rr}})^{-1} \partial_{r}$, and the extrinsic curvature is defined as $K_a^b=\nabla_a n^b$, where $a,b$, are indices of the hypersurface coordinate system. The discontinuities in the extrinsic curvature as a function of the surface  energy density $\eta$ and the surface tension $\sigma$ are \citep{mazur04}:
\begin{eqnarray}
&&[K_t^t]=\left[\frac{(\sqrt{g_{rr}})^{-1}}{2g_{tt}}\frac{dg_{tt}}{dr}\right]=\frac{4 \pi G}{c^4}(\eta+2\sigma)\,, 
\label{Kcond}  \\
&&[K_\theta^\theta]=[K_\phi^\phi]=\left[\frac{(\sqrt{g_{rr}})^{-1}}{r}\right]=-\frac{4\pi G}{c^4}\eta. \nonumber 
\end{eqnarray}
Recall that $\sigma$, as defined in this paper, has contrary sign to its usual definition. The first condition given by Eq. (\ref{Kcond}) is related to  Eq. (\ref{sigma2}). An assumed jump in the radial pressure gives a surface tension through Eq. (\ref{sigma2}), while Eq. (\ref{Kcond}) gives a jump in the first derivative of $g_{tt}$ (there is no jump in $g_{rr}$, since it is assumed that $\eta = 0$). Conversely a jump in the first derivative of $g_{tt}$ is proportional to a surface tension, and this could be calculated after obtaining the numeric solution (see the next section). 
The equation (\ref{sigma2}) is  valid at the shell even when the isotropic fermion matter is included. In this case, the stress-energy tensor as a distribution valued tensor is, 
\begin{equation}
T_{\alpha \beta}=\Theta(l) T^+_{\alpha \beta}+\Theta(-l) T^-_{\alpha \beta}+\delta(l) S_{\alpha \beta}\,,
\end{equation}
where the symbols have the usual meaning, and $l$ denotes the proper distance along geodesics from the hypersurface \citep[see][for definitions]{poisson}. The surface stress-energy tensor is:
\begin{equation}
S_{a b}=\eta\,u_a u_b- \sigma \,(h_{ab}+u_a u_b)\,,
\end{equation}
in the present case is enough to set $\eta=0$, and to choose $\sigma$ to cancel the term
$-c^4 \left( [K_{ab}]-[K] h_{ab}\right)/8 \pi$ (the ``left hand side" of Einstein equations). 
Its components were given in  Eq. (\ref{Kcond}). This is the standard procedure for dealing with thin shells in general relativity \cite{poisson}. 

A word of caution is worth mentioning. It is possible to derive an expression, from equations (\ref{DETOV}) and (\ref{eos1}), for the tangential dark energy pressure at the shell, analogous to that obtained by \cite{israel3}:
\begin{equation}
\label{IsP}
P_{t}=-\rho_{de}\,c^2\, \theta(a-r)+\frac{1}{2}\, \rho_{de}\,c^2\, a \,\delta(r-a)\,,
\end{equation}
where $a$ is the radius of the shell. 
They pointed out that this expression becomes singular at the event horizon. In fact, as they showed, a change in variables gives $\delta(r-a)=[g_{rr}(a)]^{1/2}\,\delta(s)$, in terms of the proper distance $s$ from the shell. So, if $r=a$ is light-like the pressure becomes singular, ``even when considered as a distribution" \citep{israel3}. This means that the thin shell of phase transition cannot be located exactly at the event horizon, but can be located elsewhere. Although this conclusion was originally obtained by a precursor idea of the dark energy stellar models, it is of course, applicable to the models studied here. 

\begin{figure}[h]
\begin{center} 
\includegraphics[width=8cm]{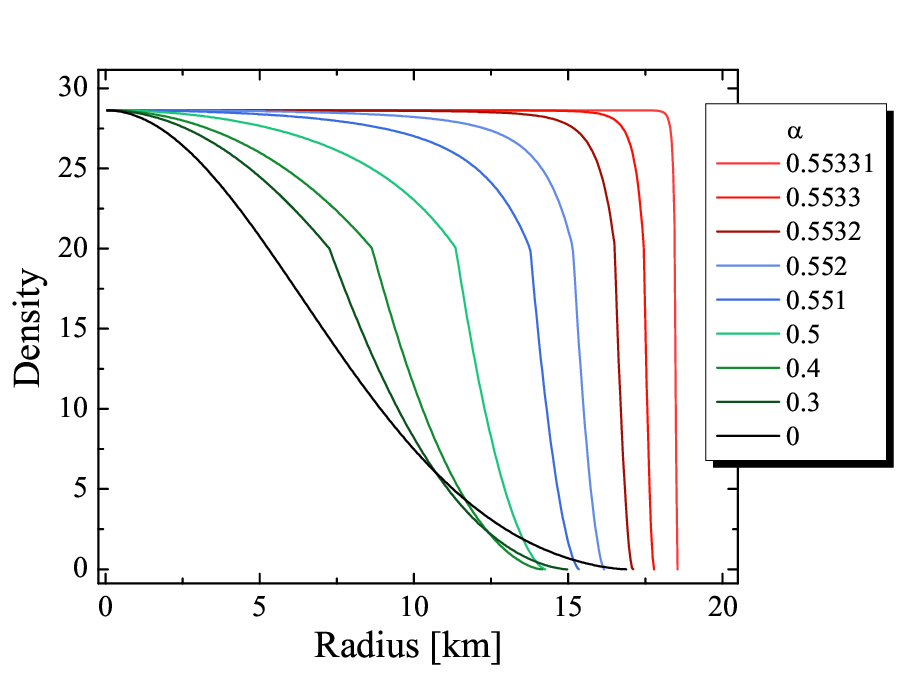}
\end{center}
\caption{Mass density profile for each dark energy star. The mass density is not smooth at the phase transition shell. The density is measured in [${\rm g\,cm}^{-3}$] and normalized by the factor $10^{13}$.}
\label{stab1}
\end{figure}

\section{\label{numeric}Numeric algorithm}
 The discontinuity in the dark pressure is managed as follows:
 the radial derivative of the total pressure in  Eq. (\ref{TOV1}) is split in two terms,
\begin{equation}
\frac{d\hat{P}}{d r}=\lim_{\delta r \rightarrow 0} \left( \frac{\delta P_r}{\delta r}+ \frac{\delta P}{\delta r} \right), \label{split}
\end{equation}
taking the limit with ${\delta r}$ tending to zero. The difficulty to integrate the $\Lambda$-TOV equations over the whole star is that the first term on the right side of Eq. (\ref{split}) diverges at the shell, i.e.: \mbox{$\lim_{\delta r \rightarrow 0} \frac{\delta P_r}{\delta r}|_{r=a}\rightarrow \infty$}. This term is associated with the second term on the right side of  Eq. (\ref{IsP}).  The solution, proposed here, is to cancel that term with an appropriate amount of anisotropy  on the right side of Eq. (\ref{TOV1}).  As explained, this is equivalent to set a finite surface tension at the shell (see Eq. \ref{sigma}). 

Now the set of equations to be integrated are equations (\ref{grr}), (\ref{totalmass}), and  (\ref{TOV1}) (without the discontinuity), complemented with the equations of state described above.  The boundary condition for the Eq. (\ref{TOV1}) is $\hat{P}=0$ at the surface of the star, and the initial condition, a prescribed mass density at the center. 

Briefly, the algorithm works as follows: first, it integrates the equation
\begin{eqnarray}
&&\frac{d{P}}{dr}=-(\delta c^{2}+P)\frac{\bigl(\frac{mG}{c^{2}}+\frac{4\pi G}{%
c^{4}}\hat{P} r^{3}\bigr)}{r\bigl(r-\frac{2mG}{c^{2}}\bigr)}\,,     \label{TOVm}
\end{eqnarray}
from the center of the star, with a given central mass density.
Next, it updates the values of $\hat{P}, \,P, \, \rho,\,\delta,\,\rho _{de}$, and $m$. If the mass density is above zero, it goes a radial step further and repeat all. If it is zero the code ends.
Of course, in the practice $\delta r$ is never zero, but equal to the radial step of the integrator with $\delta r \ll a$. It has been checked that the results are robust respect to variations in $\delta r$. 

 Note that the Eq. (\ref{TOVm}) is different from the Eq. (\ref{TOV1}). The term 
${d P_r}/{d r}$ was canceled by a surface tension $\sigma$ so that
$\delta P_r=2 \sigma/ r$.

Equations (\ref{grr}), (\ref{totalmass}), and  (\ref{TOV1}) are integrated with a Runge-Kutta algorithm of the fourth order.
The employed code HE05v2 is an update of the previous code HE05v1 \citep[see][for more details]{ghezzi}.
\begin{figure}[h]
\begin{center} 
\includegraphics[width=8cm]{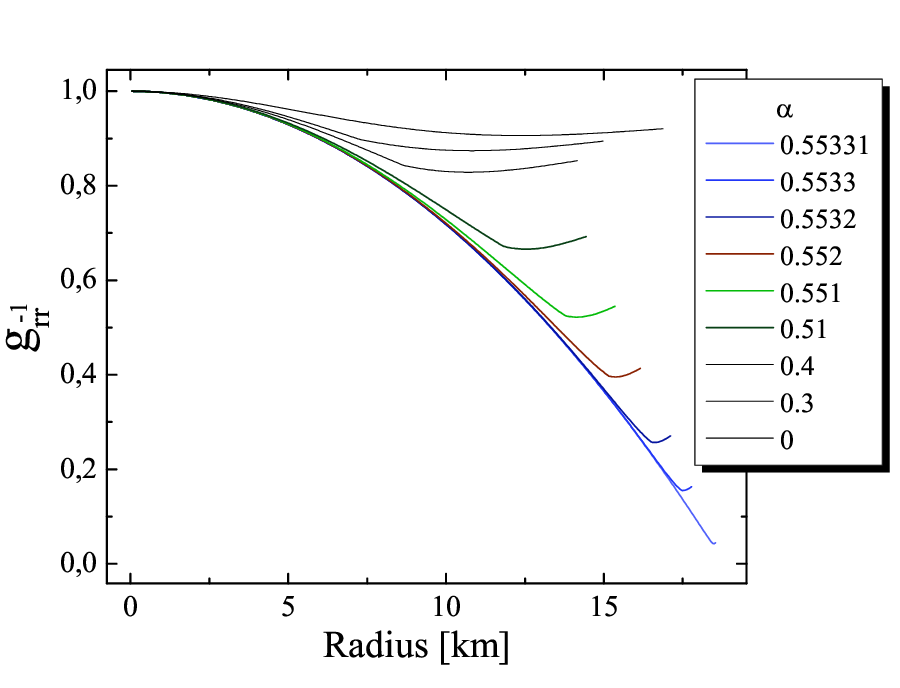} 
\end{center}
\caption{Coefficient $g_{rr}^{-1}=1-{2\,m(r)\,G}/{r\,c^{2}}$, for dark energy stars with different values
of the parameter $\alpha$. At the surface of the star $g_{rr}^{-1}=g_{tt}$.}
\label{gamma}
\end{figure}

\subsection{\label{caveats}Caveats}
Although the dark energy pressure derivative was canceled, it is not granted that the solutions will be smooth. The solving procedure described above eliminates the delta Dirac distribution from equation (\ref{TOV1}), but smoothness was not imposed anywhere in the remaining equations. As it will be shown, there are solutions which are not smooth at the transition shell. It is possible to integrate the non-smooth solutions performing the calculation at zero order precision on the shell, and at fourth order in the rest of the star. 

Another caveat is that the denominator of the Eq. (\ref{TOV1}) is  zero at the event horizon, and the equation becomes singular. A larger numeric resolution is needed as the compactness approaches one.  This problem is inherent to the field equations, and of course, it is independent of the position of the transition shell. However, it also includes the issue pointed out above: the transition shell cannot be located at the event horizon.

\begin{figure}[h]
\begin{center} 
\includegraphics[width=8cm]{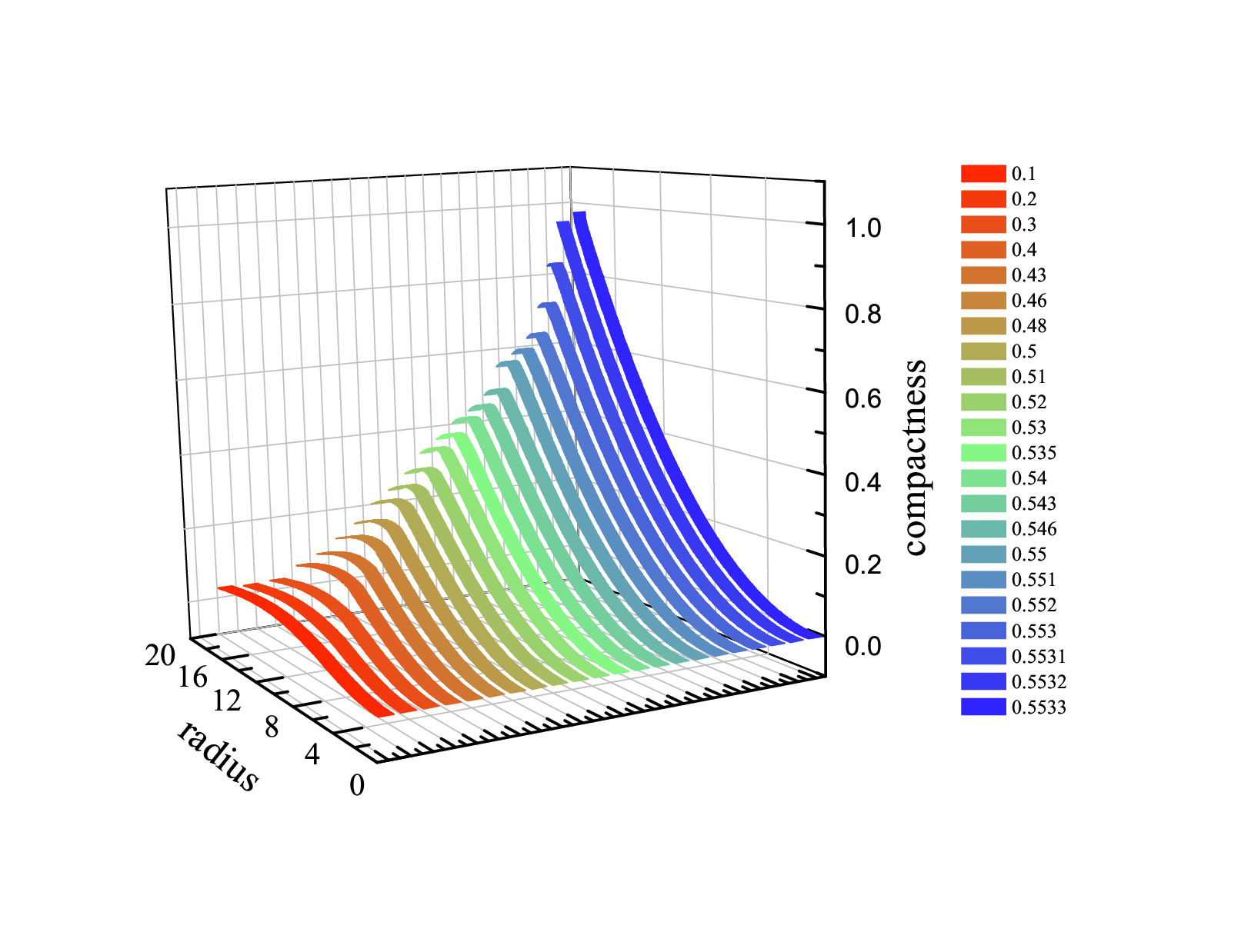} 
\end{center}
\caption{The compactness function: $r_{sch}/r$, is plotted for several dark energy stars.  As the parameter $\alpha$ is increased, the compactness at the surface tends to one, and the model approaches a regular black hole.}
\label{compact}
\end{figure}

\section{\label{disc}Discussions}
 A family of solutions is obtained, depending on $\alpha$, for each stellar central density.
In order to analyze the structure of the stars,  a model with central density: \mbox{$\rho=2.86 \times 10^{14}\,\,{\rm g\,cm}^{-3}$}, is singled out. 
The results of the calculations are shown in Figures \ref{stab2}-\ref{compact}. 
 Figures \ref{stab2}, \ref{stab1}, and \ref{gamma}, show the stellar structure of the set of dark energy stars. These stars are in the stable branch of the solutions (see below).  The different solutions were obtained varying the coupling parameter from $\alpha=0$ to $\alpha=0.5533$. 

The summary of  possible solutions is given in  Table \ref{table}, presented together with some well known solutions. 
 The type (f) solutions shown in the Table \ref{table}, are obtained setting $\alpha=0$, and  correspond to the typical Tolman-Oppenheimer-Volkoff neutron star model. 

Black hole solutions are obtained with $\alpha \ge \alpha_c=0.5533$. The critical value $\alpha_c$ is obtained by interpolation and cannot be calculated exactly \footnote{In fact the interpolated value is $\alpha_c=0.553317$, with a precision of the order of $10$ cm in the radial step.}, as was explained above (see Sec. \ref{anisot} and \ref{numeric}). The black hole solutions are characterized by a surface radius smaller than or equal to the Schwarzschild radius, or a  larger or equal to one compactness. These are indicated as type (a) and (b) solutions in Table \ref{table}. The critical $\alpha_c$ parameter depends on the particular family of solutions considered, i.e.: on the central mass density.  

 Figure \ref{stab2} shows the pressure profile for each dark energy star of the family. The pressure tends to a negative constant inside the dark energy star as the critical parameter is approached. The pressure is positive and decreasing with radius outside the thin shell.
Figure \ref{stab1} shows the mass density for each dark energy star.  The density tends to a constant value, inside the shell, as the critical parameter is approached ($\alpha \rightarrow \alpha_c$). The  density profiles are non smooth for this family, because there is a surface tension at the thin shell of phase transition (see Eq. \ref{Kcond}). This is expected in general, since the matter density need not be smooth across the shell.

 In general, the total pressure (effective equation of state) can be written as: 
\begin{equation}
\hat{P}=-w\,\rho\, c^2\,,
\end{equation}
with 
\begin{equation}
w=\left(\alpha-\frac{P}{\rho c^2}\right)\,.
\label{w}
\end{equation}
The value of $w$ is bounded from above by $\alpha$, and it depends on $(\alpha, \rho)$. For the analyzed model $w=0.5287$ for $\alpha=\alpha_c$. Thus, the total pressure is  $\hat{P}>-\,\rho\, c^2$, for the considered set of dark energy stars. 
Every family of solutions contains at least a black hole, i.e.: the critical parameter $\alpha_c$ is finite for all families of solutions. This was obtained by inspection of the numerical solutions. 

For a given value of the central density (for a given family of solutions), there exists 
an $\alpha=\alpha_c'$,  so that the effective equation of state is of the cosmological constant type at the center of the star, i.e.: $\hat{P}=-\rho\,c^2$ (do not confuse $\alpha_c'$ with $\alpha_c$). If $\alpha_c > \alpha_c'$ the sequence of solutions, as function of $\alpha$, converge on a dark energy star  of this kind (type c) before converging on a black hole. On the contrary, if $\alpha_c \le \alpha_c'$, the sequence converges first on a black hole.

\begin{table*}
\caption{\label{table}Summary of solutions.}
\footnotesize\rm
\begin{tabular*}{\textwidth}{@{}l*{15}{@{\extracolsep{0pt plus12pt}}l}}
\tableline
type&$\,$ w &$\, \alpha$&matter EOS& $\,$layers $\,$&$\,$ Anisotropy $\,$ & converges to\\
\tableline
\verb"a"&$\le1\,$&$\ge \alpha_c \,(\alpha \le \alpha_c')$& $\,\,$neutrons  $\,\,$ & $\,\,\,\,$3 & $\,\,\,\,$Yes &Black Hole\\

\verb"b"&$>1\,$&$\ge \alpha_c \,(\alpha > \alpha_c')$& $\,\,$neutrons  $\,\,$ & $\,\,\,\,$3 & $\,\,\,\,$Yes &Black Hole\\

\verb"c"&$1\,$&$=\alpha_c'\,(\alpha_c' < \alpha_c)$& $\,\,$neutrons $\,\,$ & $\,\,\,\,$3 & $\,\,\,\,$Yes & Gravastar, cosmological const. as $r \rightarrow 0$\\

\verb"d"&$\le1\,$&$<min(\alpha_c',\alpha_c)$& $\,\,$neutrons  $\,\,$ & $\,\,\,\,$3 & $\,\,\,\,$Yes & Gravastar\\

\verb"e"&$>1\,$&$\alpha_c' < \alpha < \alpha_c$& $\,\,$neutrons  $\,\,$ & $\,\,\,\,$3 & $\,\,\,\,$Yes & Gravastar\\

\verb"f"&$\,$-&  $0$& $\,\,$neutrons $\,$& $\,\,\,\,$0 & $\,\,\,\,$No & TOV star \citep{oppi}\\

\verb"-"&$\,$-& -&  $ $\,\,\,$ P=\rho /3$ $\,$& $\,\,\,\,$0 & $\,\,\,\,$Yes & anisotropic star  \citep{bowers}\\

\verb"-"&$1\,$&-&$\,$ $P=const.$ $\,$& $\,\,\,\,$3 & $\,\,\,\,$No & DE star \citep{chapline01}\\

\verb"-"&$1\,$&-&$\,$ $P=const.$  $\,$& $\,\,\,\,$5 & $\,\,\,\,$Yes &  Gravastar \citep{mazur01}\\

\tableline
\end{tabular*}
\end{table*}

\begin{figure}[h]
\begin{center}  
\includegraphics[width=8cm]{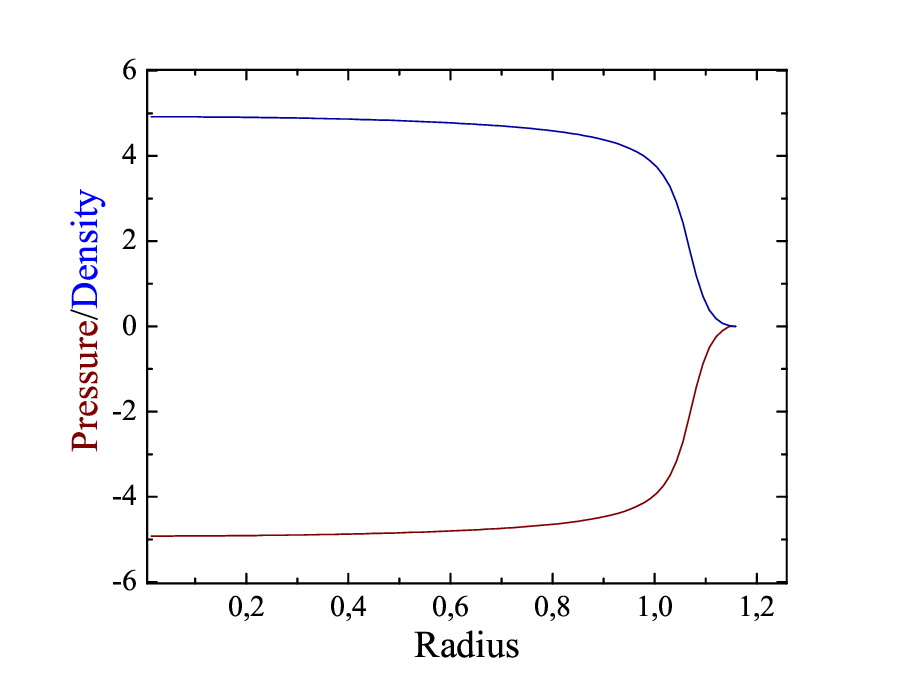}
\end{center}
\caption{Detail of the pressure and density profile with  $\alpha=\alpha_c'$ and with central density $\rho=4.92 \times 10^{16}\,\,{\rm g\,cm}^{-3}$. In this case the magnitude of the total radial pressure at  the center is: $\hat{P}=-\rho\,c^2$.   The physical units of the density and pressure ($\times c^{2}$) are: [$1e16$] gr cm$^{-3}$,  the scale is linear. }
\label{pressdens}
\end{figure}

  The critical parameter $\alpha_c'$ can be obtained exactly from the neutron gas equation of state. For example: the Fermi parameter is $x=(\rho/\rho_0)^{1/3}=0.3604$ for the solutions considered here (with central density $\rho=2.86 \times 10^{14} \, {\rm g \, cm^{-3}}$).   With $w=1$, using  Eqs. (\ref{w}) and (\ref{eosn}), is obtained:
\begin{equation}
\alpha_c'=1+\frac{3}{8} \biggr[x\, \sqrt{1+x^2}\bigl(\frac{2 x^2}{3}-1\bigr) +
\mathrm{ln} \bigl(x+\sqrt{1+x^2}\bigr) \biggr]/x^3\,,
\end{equation}
 that gives $\alpha_c'=1.0248$. The same calculation can be done for any other central density.
The value of $\alpha_c$ must be obtained numerically, i.e.: it cannot be obtained without knowledge of the complete sequence of solutions. In the present case $\alpha_c < \alpha_c'$ so, the solutions converge on a black hole before reaching a solution with an effective cosmological constant equation of state, at the center of the star.
The parameter space was swept looking for another family of solutions that contains a type (c) gravastar. 
 This is not a trivial task, because $w$ must be calculated for $\alpha=\alpha_c$, while $\alpha_c$ vary between the different families. So, a complete set of simulations had to be performed for each central density. It was found that  $w(\alpha_c)$ increases with the central density of the models. In particular, a gravastar model of type (c) was found with a central density of $\rho=4.92 \times 10^{16}\,\,{\rm g\,cm}^{-3}$. In this case, the critical parameter for the type (c) gravastar is $\alpha_c'=1.416$, while the critical parameter for a black hole solution is $\alpha_c=1.5192$. The compactness of the gravastar is $0.9079$. The structure of this star is shown in Fig. \ref{pressdens}.
Unfortunately, this model is in the unstable branch of the solutions. Of course, as explained above, it is possible to find solutions -stable or unstable- with larger compactness, but with $w \ne 1$.    

There is another type of solutions which are closer to the MM gravastar model:
  the solutions with $\alpha \rightarrow \alpha_c$, which have a  density profile  nearly constant in a large portion of the stellar interior (see Figs. \ref{stab2} and \ref{stab1}). The models with  $\alpha \rightarrow \alpha_c$ -but not reaching $\alpha_c$- have a large compactness and are not contained in a black hole. In order to have an inner de Sitter metric, the matter
density must be exactly zero at the core; this can be checked from Eqs. (\ref{grr}), (\ref{totalmass}),  and (\ref{metrica0b}).
 However, the numerical limit of the models as $\rho,\,\rho_c \rightarrow 0$  do not converge to the MM model: the density and pressure profiles keep their shape roughly, but the radius of the dark energy stars increase continuously as the density is lowered. It means that the radius of the stars tends to infinity as the central density tends to zero. The limit spacetime in this case is Minkowski (vacuum). 
But there is another way around to obtain an MM gravastar, with an inner de Sitter metric, as the limit of the numerical model: it is to change the order of the limits. First kept fixed $\rho_{de}=$constant  in the obtained solution for $\alpha \rightarrow \alpha_c$. This is equivalent to leaving the dark energy and shell radius frozen. After that, take the limit $\rho \rightarrow 0$, to obtain an inner de Sitter spacetime. 
In the process the inner mass will be overestimated, so the result is a perturbed MM gravastar on the stable branch. 
These solutions were not included in the Table \ref{table}, but can be derived from type (d) or (e) solutions. 
It is not possible to know how this gravastar will evolve with the present methods, so this issue will not be further discussed here.
 
\smallskip
\subsection{Energy conditions}
The dark energy is defined as a fluid that violates the strong energy condition (SEC).
 A distinction of all possible kind of fluids interior to a gravastar were classified in \cite{chan}, according to where the fluid violates the SEC condition, or some of the null energy conditions (NEC).

 Respect to an orthonormal basis, the SEC condition at the shell reads: $\eta c^2-2\sigma \ge 0$ and the NEC: $\eta c^2-\sigma\ge0$. In the present case, as $\eta=0$, and $\sigma \ge 0$, the SEC and NEC conditions are violated at the shell. So, the anisotropic thin shell is made of ``repulsive phantom energy", according to the \cite{chan} classification.

On the other hand, the possibilities for the inner fluid are more exotic. Locally the SEC condition for the inner fluid is: $(\delta+\rho_{de})c^2+3\hat{P}\ge0$ (recall that the fluid is anisotropic only at the shell), while if the NEC condition is satisfied: $(\delta+\rho_{de}) c^2+\hat{P}\ge0$.  
All the solutions found satisfies the weak energy condition (WEC) and the NEC.

 The solution (b) (or (e)) is a black hole (or a gravastar) with  dark energy in its interior. The dark energy violates the SEC. Thus, the formation of an inner singularity is avoided, because SEC violation produce a repulsive effect on geodesics (Raychaudhuri theorem). The solutions of type (a) (or (c)-(d)) are black holes (or gravastars) divided in two subsets according to the energy conditions. There is a subset which do not violate the SEC: this is an attractive normal fluid; while  the second subset  has a repulsive dark energy core.

\smallskip
\subsection{Compactness}
The compactness is defined as the quotient between the Schwarszchild radius and the radius of a sphere. It is a function of the radius and the parameter $\alpha$.  Figure \ref{compact} shows the compactness function for the considered family. It can be seen that the compactness of the star surface tends to one as $\alpha \rightarrow \alpha_c$.
  If $\alpha \ge \alpha_c$ the compactness is larger  or equal to one,  and the shell is contained within a black hole.   The solutions with $\alpha \ge \alpha_c$ are black holes, indicated as type (a) or (b) in Table \ref{table}. 

\smallskip
\subsection{Radial stability}
\begin{figure}[h]
\includegraphics[width=8cm]{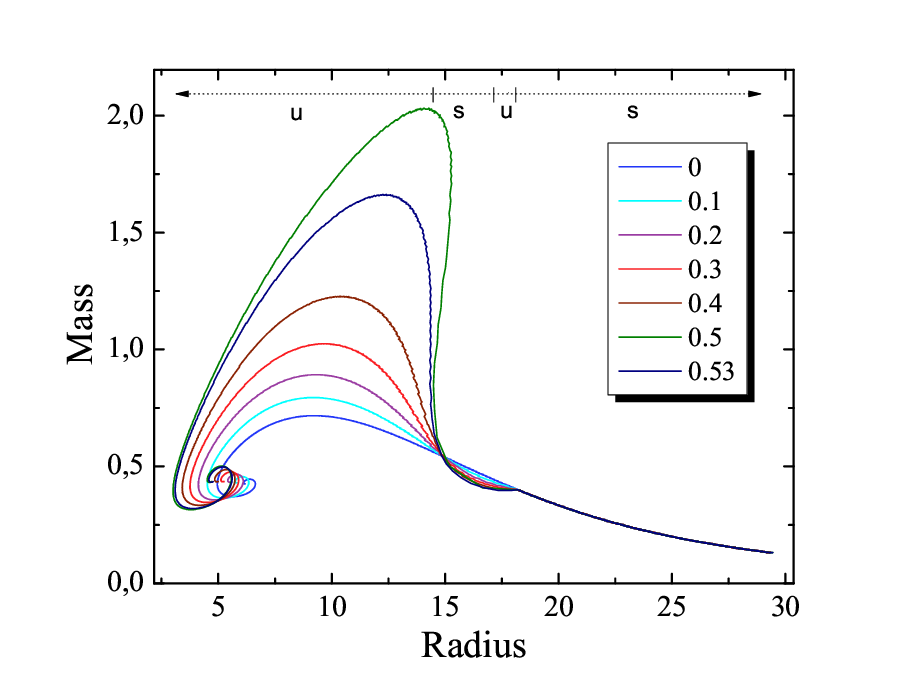}
\caption{Mass radius relation for a set of solutions.  The curve with $\alpha = 0$  corresponds to normal TOV neutron stars. }
\label{mass1}
\end{figure}

 Figure \ref{mass1} shows the mass-radius relations for dark energy stars, with a coupling parameter that varies from $\alpha=0$ to $\alpha=0.53$.  
There are two maxima in the mass-radius curve. The first one is at the density of phase transition, where the curve is non smooth. The second maxima is due to the matter equation of state, like in the usual TOV neutron stars. There is a valley in the curves because at low masses the repulsive effect of the dark energy is more important than its weight. However, at larger masses the weight overcomes the negative pressure effect and the masses of the dark energy stars become larger than fiducial TOV stars with the same radius.

The dark energy star of the family is initially (at $\alpha=0$) on the stable region near the first maximum,  but its position on the mass-radius curve is displaced to the left, to lower radius, as  $\alpha$ increases.

 The criteria  $dM/d\rho>1$ ($dM/d\rho<1$) is a necessary condition for hydrodynamic stability (instability), respectively, but it is not a sufficient condition \cite{glendenning}. However, it is possible to apply a Wheeler's $M(R)$ analysis to the mass-radius relation.
The method was devised by Wheeler, and the precise conditions under which it is valid, were given by Thorne \citep[see][]{thorne, thorne2,thorne3}. It is useful, in the present case, to obtain the number of unstable radial modes for each dark energy star in hydrostatic equilibrium without calculating the eigenfunctions and eigenfrequencies of the radial perturbations.  The condition for the applicability of the method is satisfied: the microscopic equation of state represents cold catalyzed matter, i.e.: no changes in nuclear composition are considered \footnote{For a more realistic equation of state the Wheeler $M(R)$ analysis is applicable if the adiabatic index satisfies a certain criteria \citep{thorne}.} The method is applied as described in \cite{thorne3}: the graph of the mass versus radius, is parameterized by the central density of the configuration (for a particular $\alpha$ value).  As we move in the direction of increasing density all configurations remain stable until the first maximum value of M is reached. At the first extremal point, the fundamental mode of radial oscillation becomes unstable.  At the second extremal point, another mode of radial oscillation changes stability. The direction of stability change depends on the shape of the $M(R)$ curve at that extremal point. If the curve bends clockwise, with increasing density, then one previous unstable mode becomes stable. On the contrary, if the curve bends counterclockwise, then one previous stable mode becomes unstable.  The same criterion is applied for each succeeding extremal point \footnote{This description of the Wheeler analysis is based on the paper of \cite{thorne}. Another clear exposition is given in \cite{thorne2}.}.

It is generally assumed that a cold star can be dynamically unstable against nonradial perturbations only if its lowest radial mode is also unstable \citep[see][]{thorne2}. Thus the radial stability analysis reveals the absolute stability and instability of the dark energy stars.
According to this analysis, the stability and instability regions for $\alpha=0.53$ are indicated in the Fig. \ref{mass1}, with the letters ``s" and ``u" respectively.
 There appears an unstable band between two stable ones, starting at the density of phase transition and ending at the bottom of the valley. This is analogous to the unstable band between the white dwarfs and the neutron stars \cite[see][]{thorne}.

\section{\label{remarks}Final Remarks}

In this work, the $\Lambda$-TOV anisotropic equations were solved for  compact objects composed by a Fermi gas coupled to dark energy. The dark energy traces the matter distribution at high energy density, with a strength depending on the parameter $\alpha$. 
 The main difference is that, in the original dark energy stellar model, the core of the star is de Sitter and the shell is located close, or replacing, the event horizon. In the solutions found in this paper, the position of the shell is calculated from the model and not imposed as a boundary condition.
Moreover, unlike other gravastar models, the DE stars considered here contain matter in its core (an exception is the model of \cite{mbonye}). 

A simple numerical method was devised to integrate the stellar structure equations from the center of the star to its surface, including the shell.  The compactness of the stars, their structure and mass-radius relations were calculated for comparison with the well known theoretical TOV model. 

It is found that the dark energy plus anisotropic stress helps to stabilize a neutron star, but there is yet a maximum mass for stable DE stars.

It has been shown that the effective equation of state is: $\hat{P}=-w\,\rho\,c^2$, with  $w=1$, $w<1$, or $w>1$, depending on the density and the assumed value of $\alpha$. 
There is a critical parameter $\alpha_c'$ so that the total pressure, at the center of the star, is of the cosmological constant type $\hat{P}=-\rho\,c^2$. 

There exists, also, a critical parameter $\alpha_c$ which separates dark energy stars from   regular black hole solutions. 

  The mass-radius curves for the dark energy stars have a maximum mass, signaling the unavoidability of black hole formation for large enough mass. It is unlikely that some other EOS for the matter will avoid the BH formation. The only possibility left is a different interaction law between dark energy and matter, for which we have no clues at present. This issue must be further studied.    

One of the characteristic features of the mass-radius relation for the dark energy stars is that there appears an instability gap between the two stable bands, which do not exist for TOV neutron stars.  It seems probable that these unstable stars migrate to the stable branch at larger radius. The detection of this gap could imply a large interaction between dark energy and matter, and could be a clear signature of the validity of the model.  However, its negative detection does not rule out the existence of dark energy stars. In this case, their identification through the mass-radius curve could be more difficult.

\nocite{*}
\bibliographystyle{spr-mp-nameyear-cnd}
\bibliography{biblio-u1}

\end{document}